\documentclass[conference,a4paper]{IEEEtran}
\IEEEoverridecommandlockouts
\usepackage{cite}
\usepackage{amsmath,amssymb,amsfonts}
\usepackage{algorithm}
\usepackage{algpseudocode}
\usepackage{graphicx}
\usepackage{textcomp}
\usepackage{blindtext}
\usepackage{subfigure}
\usepackage{float} 
\usepackage{subfigure}
\usepackage{xcolor}
\usepackage{diagbox}
\usepackage{slashbox}
\usepackage{wrapfig}
\usepackage{makecell}
\usepackage[left=1.7cm,right=1.7cm,top=1.9cm]{geometry}
\def\BibTeX{{\rm B\kern-.05em{\sc i\kern-.025em b}\kern-.08em T\kern-.1667em\lower.7ex\hbox{E}\kern-.125emX}}

\makeatletter
\begin{document}
	\title{Investigating the Impact of Variables on Handover Performance in 5G Ultra-Dense Networks}
	\IEEEpeerreviewmaketitle
	\author{\IEEEauthorblockN{Donglin Wang\IEEEauthorrefmark{2}, Anjie Qiu\IEEEauthorrefmark{2}, Qiuheng Zhou\IEEEauthorrefmark{1}, Sanket Partani\IEEEauthorrefmark{2} and Hans D. Schotten\IEEEauthorrefmark{2}\IEEEauthorrefmark{1}}
		\IEEEauthorblockA{\textit{\IEEEauthorrefmark{2}University of Kaiserslautern, Kaiserslautern, Germany} \\
		$\{$dwang,qiu,partani,schotten$\}$@eit.uni-kl.de \\}
		\IEEEauthorblockA{\textit{\IEEEauthorrefmark{1}German Research Center for Artificial Intelligence (DFKI GmbH), Kaiserslautern, Germany} \\
		$\{$qiuheng.zhou,schotten$\}$@dfki.de}
	}
	\maketitle
	
\begin{abstract}
The advent of 5G New Radio (NR) technology has revolutionized the landscape of wireless communication, offering various enhancements such as elevated system capacity, improved spectrum efficiency, and higher data transmission rates. To achieve these benefits, 5G has implemented the Ultra-Dense Network (UDN) architecture, characterized by the deployment of numerous small general Node B (gNB) units. While this approach boosts system capacity and frequency reuse, it also raises concerns such as increased signal interference, longer handover times, and higher handover failure rates. To address these challenges, the critical factor of Time to Trigger (TTT) in handover management must be accurately determined. Furthermore, the density of gNBs has a significant impact on handover performance. This study provides a comprehensive analysis of 5G handover management. Through the development and utilization of a downlink system-level simulator, the effects of various TTT values and gNB densities on 5G handover were evaluated, taking into consideration the movement of Traffic Users (TUs) with varying velocities. Simulation results showed that the handover performance can be optimized by adjusting the TTT under different gNB densities, providing valuable insights into the proper selection of TTT, UDN, and TU velocity to enhance 5G handover performance.

\end{abstract}


\section{introduction}
The exponential growth of mobile data globally, particularly in connected vehicle applications, has fueled the development of 5G NR technology as outlined in 3rd Generation Partnership Project (3GPP) releases 15 and 16 \cite{3gpprel15}\cite{3gpprel16}. One of the defining features of 5G is the deployment of the UDN, aimed at meeting the demand for high data traffic. The IMT-2020 group recognizes UDN as a critical component of 5G core technologies, which significantly improves spectrum efficiency and system capacity by increasing the coverage of 5G gNBs while reducing the number of served TUs per gNB \cite{IMT2020(5G)PG}. However, the deployment of UDN in 5G is anticipated to cause longer handover times, thus reducing the overall handover performance \cite{HOLTENRsurvey2019}. The increased capacity of 5G through UDN comes at the cost of higher handover rates and increased signal overheads, highlighting the importance of effectively managing handovers in 5G networks.

Before delving into the impact of 5G gNB density on the 5G handover rate, it is essential to understand the handover procedure in both LTE and 5G NR. A comprehensive documentation survey of handover management in LTE and 5G NR is presented in \cite{HO5G2016}, highlighting the critical aspects and challenges that must be considered when developing an optimized handover scheme. In \cite{5gHO2021}, a formal analysis of 5G handover is presented, covering various aspects such as protocol testing, verification, mobile networking, and more. To date, research efforts have primarily focused on optimizing the LTE handover scheme \cite{lteho2009}\cite{ShiUDN2019}. Nevertheless, with the advent of 5G NR technology, it is imperative to assess the impact of 5G gNB density on the handover rate, which has not been thoroughly explored in previous studies.

TTT is a valuable parameter in the performance of the 5G handover rate. The handover is processed between two neighboring cells if the criterion of handover is met during TTT \cite{3gppTR36816}. Using the TTT time period can prevent excessive frequent handover events in a short time period. But a too large TTT time period may cause connection loss or bad connection quality of TU from one cell to another cell. It is necessary to detect the effect of various TTT values on the handover performance. In \cite{TTTforLTEHO}, Juwon presents a handover optimization scheme for different speeds of TUs in LTE. In this work, adjustable TTT parameters are applied. In this work \cite{TTTSONLTE2010}, the effects of TTT values on the handover performance in an LTE system are analyzed. In \cite{NguyenLTETTT2017}, a mobility robustness optimization method for handover failure reduction in LTE small-cell networks is proposed by adjusting TTT and offset parameters. The results from \cite{TTTSONLTE2010}\cite{NguyenLTETTT2017} show the handover performance is improved by applying adaptive TTT parameters. However, these works evaluated the effect of various TTT values on the handover performance for LTE handover not for 5G NR handover performance. In our work, we want to evaluate the effect of different TTT values on the 5G UDN handover performance by applying varying densities of gNB from the 5G network.

The paper is organized as follows: for section II, we are going to show what the 5G handover structure looks like, and how to make a handover-triggering decision. Section III establishes the 5G UDN simulation scenario and sets simulation parameters. Section IV analyzes the simulation results for different simulation scenarios. Lastly, in section V, the conclusion and future work plan are drawn. 

\section{5g handover process}
In this work, the 5G NR handover procedure is considered and is also developed based on LTE handover technologies. But there are some differences between LTE handover and 5G NR handover \cite{HOLTENRsurvey2019}. 5G NR handovers can be performed in a more dynamic manner, allowing for a smoother transition between cells, as well as improved performance and user experience. 5G NR handovers can also be performed at a much faster rate, with less latency and delay, compared to LTE handovers, which can improve network performance and efficiency.

In 5G NR cellular networks, the mobility mechanisms enable the TU to move within the network coverage and be served by networks. The Radio Access Mobility (RAM) of a TU can be in two states, IDLE\_MODE or CONNECTED\_MODE. The TU in IDLE\_MODE performs cell selection to receive incoming data, whereas the TU in CONNECTED\_MODE actively transmits or receives data. Handover only occurs when the TU is in the CONNECTED\_MODE and a new cell is determined to be superior to the current serving cell \cite{HO5G2016}. 
To be more specific, handovers in 5G networks can be classified into two categories: intra-layer-handover and inter-layer-handover. The distinction between the two categories lies in the serving and target networks, whether the same Radio Access Network (RAN) technologies are used.  If the same RAN technology is employed by the serving and targeting networks, an intra-layer-handover is performed. Conversely, if different RAN technologies are used, an inter-layer-handover is required, such as transitioning from 5G to LTE or vice versa. In this study, we only consider intra-layer-handover between 5G networks, as specified by 3GPP Release 16.3.0 \cite{3gpprel16.3.0}. 

\begin{figure}[htbp]
	\centering
	\includegraphics[width=\linewidth]{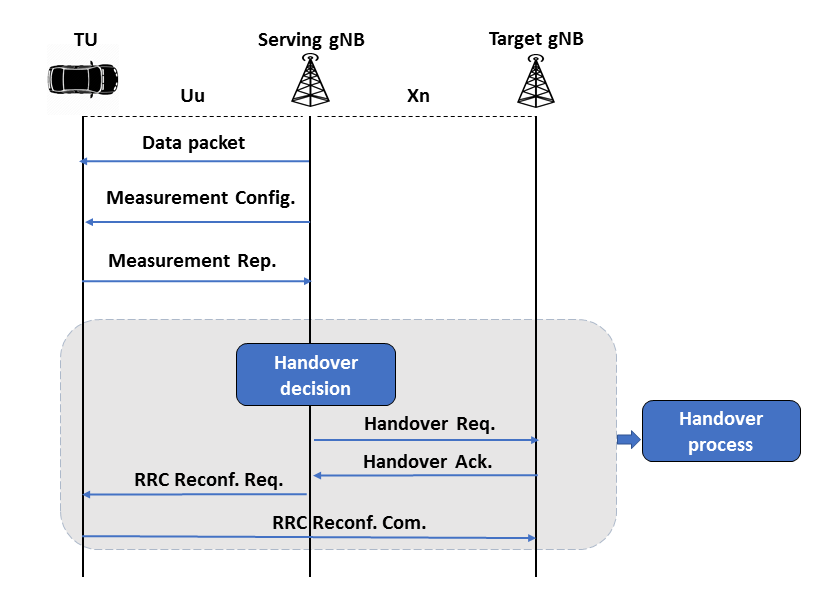}
	\caption{Simplified 5G handover procedure}
	\label{5GHO}
\end{figure}

\subsection{Overview of 5G Handover Procedure}
In Fig. \ref{5GHO}, a simplified illustration of the 5G handover procedure is presented. This procedure can be generally divided into three stages.
\subsubsection{Measurement and monitoring stage}
The TU is engaged in the data communication with the Serving gNB via the Uu interface. The Serving gNB provides the TU Measurement Configuration (Config.) information through a re-connection message. The TU performs and processes the collected measurements of the Received Signal Strengths (RSSs), and then transmits the Measurement Report (Rep.) back to the Serving gNB at designated intervals of 10 milliseconds.

\begin{figure}[htbp]
	\centering
	\includegraphics[width=.9\linewidth]{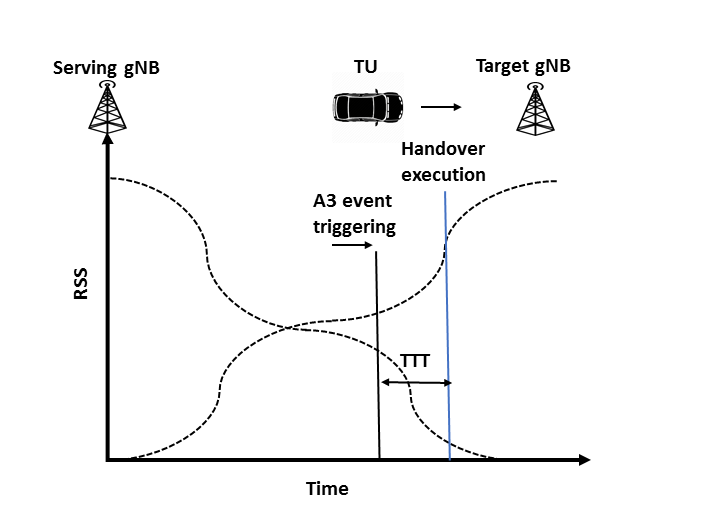}
	\caption{A3 event triggering}
	\label{A3}
\end{figure} 

\subsubsection{Handover Decision Making}

The procedure of handover decision-making in 5G networks involves the assessment of RSS of both the Serving and Target gNBs. The assessment is triggered by the A3 event \cite{3gppTR36816}, which requires that the RSS of the Target gNB is better than that of the Serving gNB, along with a Hysteresis (Hys) margin of 3dB as shown in Fig. \ref{A3}. The A3 event must be maintained for a specific time period referred to as the TTT timer. The TTT time is a crucial factor in determining the reliability and frequency of handovers in 5G networks. If the conditions are met, the Serving gNB makes the handover decision and initiates the process by sending a Handover Request (Req.) to the Target gNB via Xn interface as shown in Fig. \ref{5GHO}. 

\subsubsection{Handover Execution Process}

The actual execution of the handover occurs in this stage. The communication between the TU and the Serving gNB is served, and a new connection is established between the TU and the Target gNB. Upon successful connection to the Target gNB, the handover process is considered complete. For further details regarding the actual steps involved in the handover execution, please refer to \cite{5gHO2021}.

\subsection{Algorithm for 5G handover-triggering}
In this part, the functions for the 5G handover triggering algorithm are provided. 

Step 1: The Signal-to-Interference-plus-Noise Ratio (SINR) (in dB) for the TU is calculated at each tic, which served as an indicator of the RSS quality. It measures the strength of the desired signal compared to the unwanted interference and noise:

\begin{equation}
sinr_i=10log_{10}  \left ({\frac{P_i}{\sum_{\mkern-5mu k \neq i} P_k +N_0 }} \right)   \label{eq1}
\end{equation}

where $sinr_i$ is the SINR of the TU at one tic w.r.t current Serving gNB $i$, $P_i$ is the power received from the current Serving gNB $i$. The received power is calculated based on transmit power, pathloss (function of distance, frequency, or antenna heights), shadowing, fast-fading, and antenna gain as shown in TABLE \ref{SP}. $k$ represents other gNBs except for the current Serving gNB $i$. The number of other gNB cells is dependent on the density of gNBs and their communication ranges, both of which will be examined in the scenario configuration section. $P_k$ is the interference power from other gNBs. In addition, $N_0$ is the noise figure. 

Step 2:
In this step, we find the best SINR for the TU. The $best\_sinr_j$ represents the best SINR value from the Target gNB $j$ as shown in Eq. \ref{bestsinr}. 
\begin{equation}
best\_sinr_j =  MAX \left ( sinr_1, sinr_2, sinr_3,...sinr_x \right)   \label{bestsinr}
\end{equation}
where $sinr_1$ to $sinr_x$ are the SINR values from all other reachable gNBs. 

Step 3: 
The 5G handover logical algorithm is  described in detail in Algorithm \ref{algorithm}. The used simulation parameters are described in TABLE \ref{SP}. 

\begin{algorithm}
    \caption{5G handover triggering logical algorithm}\label{euclid}
    \hspace*{\algorithmicindent} \textbf{Input}: serving\_gnb, serving\_sinr, best\_sinr, target\_gnb, sinr\_min, avg\_sinr, best\_cio, current\_cio, ho\_hys, TTT, ho\_timer, ho\_trigger, ho\_exec\_time    \\
    \hspace*{\algorithmicindent} \textbf{Output}: ho\_times
    \begin{algorithmic}[1]
    \If {$serving\_{gnb} \neq target\_{gnb}$} 
        \If{$best\_sinr > sinr\_min \:\&\: 
            best\_sinr - avg\_sinr + best\_cio - current\_cio) > ho\_hys$}
         \State  $ho\_trigger\gets  1$
         \State  $ho\_timer  \gets  ho\_timer +1$
            \If{$ho\_timer == TTT$}
             \State $serving\_gnb \gets  target\_gnb$
             \State $ho\_exec\_time\gets  25 $
             \State $ho\_times \gets ho\_times + 1 $
             \State $ho\_trigger \gets  0$
             \State $ho\_timer   \gets  0$
            \EndIf
         \EndIf
    \EndIf
    \end{algorithmic}
    \label{algorithm}
\end{algorithm}
As demonstrated in Algorithm \ref{algorithm}, the objective is to determine the number of successful handover times in each simulation. The handover triggering logic is applied after the subsequent steps. 
1) It is ensured that the selected $target\_gnb$ is distinct from the current $serving\_gnb$, implying that they are located in disparate locations; 
2) it is to check the calculated $best\_sinr$ exceeds the pre-defined minimum SINR threshold  of $sinr\_min = -7 dB$. The handover trigger also evaluates the difference between $best\_sinr$ and $avg\_sinr$ of the TU, as well as the effect of the load balancing algorithm on the current connection $current\_cio$ and the potential connection $best\_cio$.  The difference must be greater than the handover Hys of 3 dB.
In addition, $avg\_sinr$ is the average SINR of the TU determined by taking the average of the previous 10 SINR values and recalculating at each tic in the simulation. The calculation of the average SINR can only occur once the previous $10$ values have been obtained; 3) and 4) if all the conditions are met, the $ho\_trigger$ flag is set to 1 and the $ho\_timer$ counter is incremented by one; 5) it is to check whether the $ho\_timer$ value is equal to the predefined TTT value, then the handover process is triggered; 6) it is the execution step, the value of the $serving\_gnb$ is updated to the $target\_gnb$. The $ho\_exec\_time$ is set to 25 tics. The output counter $ho\_times$ for the number of successful handovers is incremented by one, and the flags for handover triggering $ho\_trigger$ and handover timer $ho\_timer$, are reset to zero. If any of the conditions specified in the handover triggering logic are not met, the algorithm will be re-executed.

\section{Simulation scenarios}
The objective of this paper is to analyze the performance of 5G handover on variable TTTs and various UDNs. Thus, a system-level downlink simulator is implemented in Python. 
\subsection{gNB deployment}

As depicted in Fig. \ref{simscenario}, two simulation scenarios have been established where the deployment of gNB follows the Poisson Point Process (PPP) \cite{PPP2017}. In these scenarios, it's assumed that all gNBs process similar technical characteristics, including identical transmission power. In Fig. \ref{simscenario}, an example of $den\_gNB = 20$ is given which indicates 20 gNBs following PPP distribution in a 1000 m x 1000 m urban area. To evaluate the impact of UDNs on 5G handover, we have conducted simulations using $den\_gNB$ values of 10, 20, 30, 40, and 50. 
\subsection{User mobility model}
\begin{figure}
    \centering
    \subfigure[Case A]{\includegraphics[width=0.24\textwidth]{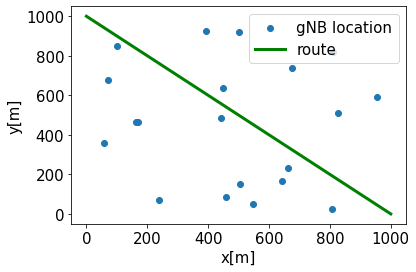}} 
    \subfigure[Case B]{\includegraphics[width=0.24\textwidth]{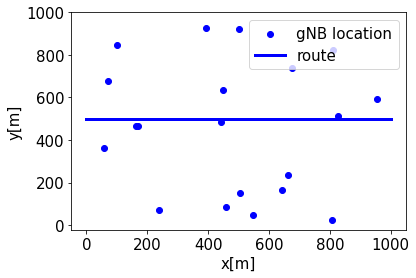}}
    \caption{Simulation scenarios with den\_gNBs (20) for different directions}
    \label{simscenario}
\end{figure}

In our simulations, we have defined two distinct TUs with varying running routes from different directions. An illustration of one of these routes is shown in Fig. \ref{simscenario}(a), where the starting point and ending point are [1000,0] and [0,1000] respectively, with a speed of 50 km/h and the direction angle ($\theta$) is 135 degrees.
Further information can be found in TABLE \ref{userroute}.  A comprehensive list of all simulation parameters, including both physical layer and system layer details, can be found in TABLE \ref{SP}.



\begin{table}[htbp]
\caption{User route models}
\begin{center}
\begin{tabular}{|p{0.9cm}|p{2cm}|p{2cm}|p{2cm}|}
 \hline
    Case A & starting point:[1000,0] & ending point:[0,1000] & direction:   $\theta = 135$\\
 \hline
    Case B & starting point:[1000,500] & ending point:[0,500] & direction:  $\theta = 180$\\
 \hline
\end{tabular}
\end{center}
\label{userroute}
\end{table}

\begin{table}[htbp]
\caption{Simulation parameters}
\begin{center}
\begin{tabular}{ |p{3cm}|p{4.8cm}|  }
 \hline
    \textbf{Parameters} &  \textbf{Description and Value} \\
 \hline
    Scenario & city in 1000 m*1000 m area \\
 \hline
    TU velocity& 10, 20, 30, 40, 50 (km/h) \\
 \hline
    den\_gNB & density of gNBs per square kilometer (10, 20, 30, 40, 50)\\
 \hline 
   $\theta$ & direction of the TU (in degree) \\
 \hline 
   TU running time & 70000 ms \\
 \hline
    gNB height & 15 m\\
 \hline
    gNB coverage& 300 m\\
 \hline
    Carrier frequency& 6 GHz\\
 \hline
    bw & bandwidths (10 MHz: 50 PRBs)\\
 \hline
    Transmit power & 30 dB \\
 \hline
    gNB antenna gain & 15 dBi\\
 \hline
    Receiver antenna gain & 0\\
 \hline
    Noise power &  -174 dBm/Hz + $10*\log_{10}$(bw) + 7\\
 \hline
    Pathloss model & pathloss = 128.1 + $37.6*\log_{10}$(Distance) \\
 \hline
    1 tic & 10 ms \\
 \hline
    serving\_gnb  & current Serving gNB location \\
  \hline
    target\_gnb  & Target gNB location \\
 \hline
    serving\_sinr & the SINR of TU from current Serving gNB \\
 \hline
    best\_sinr &  the SINR of TU from the best connection gNB \\
 \hline
    sinr\_min & the minimum SINR value to keep TU connected to gNB (-7 dB)\\
 \hline 
    avg\_sinr & average SINR of the TU w.r.t current gNB \\
 \hline
    ho\_avg\_sinr\ & the average SINR value of TU for each successful handover (dB)\\
 \hline
    best\_cio & cell individual offset (0 dB) \\
 \hline
     current\_cio & cell individual offset (0 dB) \\
 \hline
    ho\_hys & handover hysteresis threshold (3 dB) \\
 \hline
    ho\_timer & handover counter \\
 \hline
    ho\_trigger & handover trigger flag (0 or 1)\\
 \hline
    ho\_exec\_time & handover execution time (25 tics)\\ 
 \hline 
    ho\_times & handover counter for summing the total times of handover \\
 \hline
\end{tabular}
\end{center}
\label{SP}
\end{table}

\section{simulation results analysis}
The results of the simulations are analyzed to assess the impact of UDNs and TTTs on 5G NR handover times and performance. In order to obtain statistically robust results, we run each simulation 100 times and calculate the average value. The final results presented in this paper are based on these average values. 
\subsection{KPIs for measuring the handover effect}
To evaluate the performance of the 5G handover in UDN scenarios, we have employed two Key Performance Indicators (KPIs). These KPIs provide an objective measure of the handover's effectiveness and enable us to compare the results of different simulations.
\subsubsection{Handover rate}
One of the KPIs is the average number of handover times per TU after each simulation run. This metric, referred to as the handover rate, reflects the number of successful handover events. When the value of the average handover rate is less than 1, it indicates a handover failure. 
\subsubsection{Average SINR value}
The second KPI we have used is the average SINR of a TU after each simulation run. The average SINR value $ho\_avg\_sinr$ is calculated below: 

\begin{equation}
\begin{aligned}
   ho\_avg\_sinr = & MEAN (best\_sinr_{j},best\_sinr_{k},\\
   & best\_sinr_{m},..., ) \label{eq}
\end{aligned}
\end{equation}
where $best\_sinr_{j}$ is defined in Eq. \ref{bestsinr} which indicates the best SINR value of Target $gNB_j$. This metric represents the handover performance and reflects the average SINR value of a TU after every successful handover execution. The higher the average SINR, the better the handover performance.
\begin{figure*}[htbp]
	\centering
	\subfigure[Case A]{
		\includegraphics[width=0.35\textwidth]{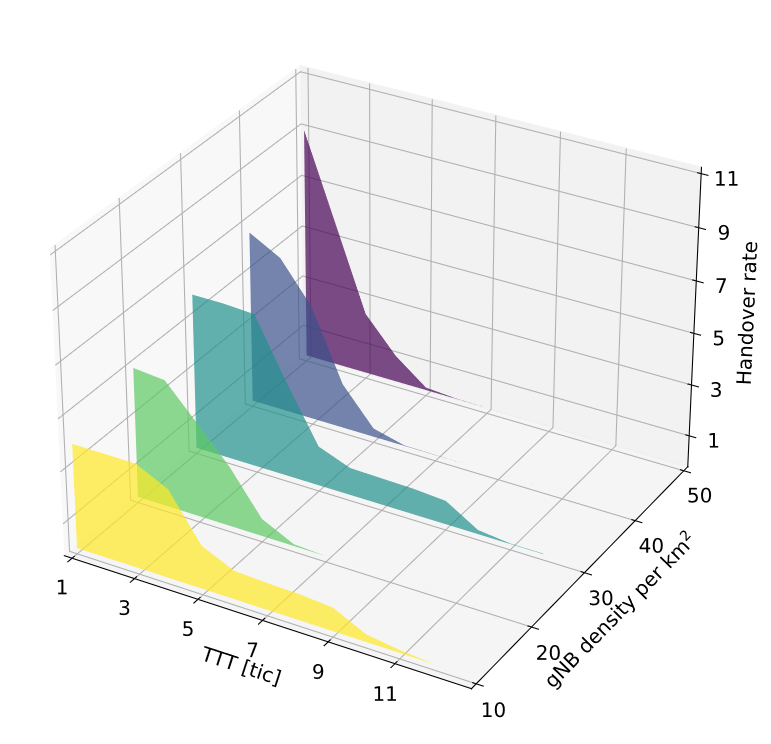}}
	\label{fig.sub.1}
	\subfigure[Case B]{
		\includegraphics[width=0.35\textwidth]{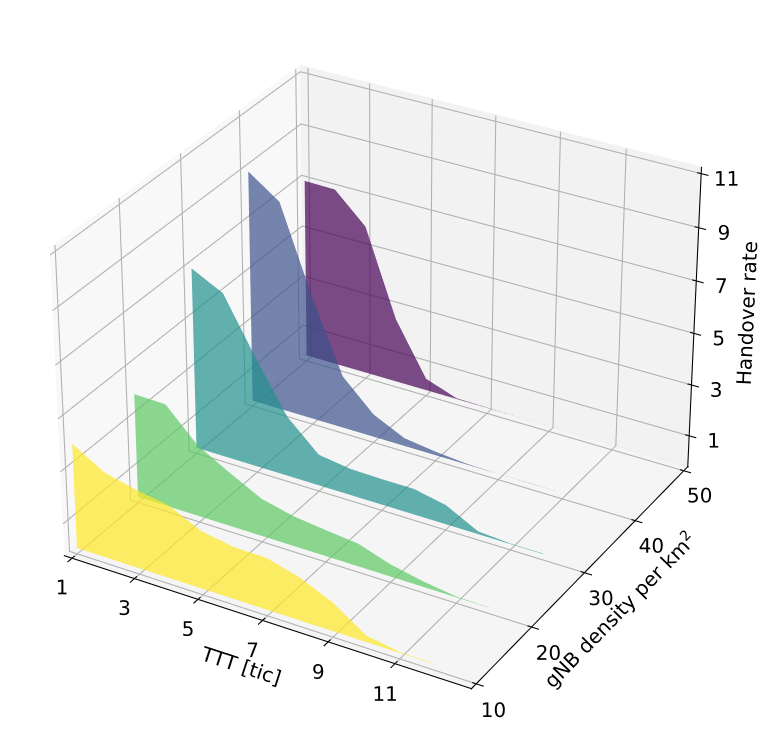}}
	\label{fig.sub.2}
    \caption{Simulation results for the two different scenarios}
    \label{resultsforscenarios}
\end{figure*}

\subsection{Effect of variable TTT values and density of UDN with a fixed velocity on handover}
To obtain a comprehensive understanding of the effect of TTTs and UDN density on handover, we conduct simulations with a range of TTT values (1 to 12 tics), UDN densities (10, 20, 30, 40, and 50 gNBs), and a fixed velocity of 50 km/h. The results of these simulations are shown in Fig. \ref{resultsforscenarios}, where 3-dimensional figures are generated for the two assumed simulation scenarios. Our results indicate that the starting location or direction of the TU does not significantly affect the final outcomes, as the simulation results for both Case A and Case B are nearly identical. This suggests that the handover mechanism in 5G networks is robust to such variations.

In Fig. \ref{resultsforscenarios}(a) case A, it is easy to find that the overall handover rate decreases with increasing TTT values when the UDN density is 10 gNBs. Specifically, the handover rate drops from 4 to 0.01 as the TTT increases from 1 tic to 12 tics. This suggests that in order to reduce the handover rate, it is necessary to increase the TTT value. However,  if the TTT value is significantly increased, such as 12 tics, the handover rate may fall below 1, resulting in handover failure. This is because a larger TTT value may cause severe degradation of SINR during the TTT period. These results highlight the importance of understanding the effect of the TTT value on the handover rate and selecting an appropriate TTT value. It is crucial to strike a balance between reducing the handover rate and avoiding handover failure.

In Fig. \ref{resultsforscenarios}(b) case B, when TTT is 1, the handover rate rises from 4 dB to 8.97 with the increasing density of gNB from 10 to 50. It means that with the increasing density of gNB, the handover rate will be significantly raised when the TTT value is not large. Ultra-dense deployment of gNB leads to redundant handovers. The impact of UDN density on the handover rate is mitigated as the TTT value increases. As the TTT surpasses 8, the density of UDN no long influences the handover rate to a significant extent. The larger the TTT for handover procedures, the weaker the effect of UDN density on handover frequency.

\begin{table}[htbp]
\caption{Average SINR of TU for Case A}
\begin{center}
\begin{tabular}{ |p{2.8cm}|p{0.5cm}|p{0.5cm}|p{0.5cm}|p{0.5cm}|p{0.5cm}|  }
\hline
\backslashbox[32mm]{\textbf{TTT}}{\textbf{ho\_avg\_sinr}}{\textbf{density}}
&\makebox{10}&\makebox{20}&\makebox{30}
&\makebox{40}&\makebox{50} \\
 \hline
    1 & 16.22 & 6.54  & 8.06  & 4.28   & 5.75\\
 \hline
    2 & 16.97 & 7.44  & 9.17  & 5.2    & 7.6\\
 \hline 
    3 & 22.54 & 10.58 & 13.33 & 8.23   & 11.49\\
 \hline 
    4 & 33.19 & 12.97 & nan   & nan    & 18.81\\
 \hline
    5 & 38.44 & nan   & nan   & nan    & nan\\
 \hline
    6 & 39.76 & nan   & nan   & nan    & nan\\
 \hline
    7 & 40.9  & nan   & nan   & nan    & nan\\
 \hline
    8 & 40.35 & nan   & nan   & nan    & nan\\
 \hline 
    9 & nan   & nan   & nan   & nan    & nan \\
 \hline
    10 & nan  & nan   & nan   & nan    & nan\\
 \hline
    11 & nan  & nan   & nan   & nan    & nan\\
 \hline
    12 & nan  & nan   & nan   & nan    & nan\\
 \hline
\end{tabular}
\end{center}
\label{caseA}
\end{table}

\subsection{Performance on handover}

Additionally, the tables present the average SINR values for two scenarios which clearly demonstrate the impact of TTT and UDN density on handover performance. The tables display the values for TTT, UDN density, and the average SINR ($ho\_avg\_sinr$) for each scenario.

In TABLE \ref{caseA} for case A, the results of the simulation indicate that in case the density of UDN is 10, the optimal TTT value for the best handover performance is 8, as it yields an average SINR of 40.35 dB. However, when the TTT value is continually increased, the handover performance deteriorates, as the TU is unable to maintain a connection with the network. As the density of UDN increases, the range of optimal TTT values becomes more restricted, and the optimal TTT value decreases. For instance, when the density of UDN is 20, the best TTT value is 4. With a density of 30 or 40, the optimal TTT value is 3. This highlights the importance of selecting the appropriate TTT value in order to maintain an optimal handover performance, especially in scenarios of high UDN density.

From TABLE \ref{caseB} case B, the simulation results obtained from case A and case B show slight variations.

For a density of 10 gNBs, the optimal TTT value is 7 with an average SINR of 30.36 dB in case B. However, as the density increases to 20, the best TTT value decreases to 6, resulting in a decrease in the average SINR to 29.46 dB. Additionally, for both cases, A and B with a TTT value of 1, the performance of TU in a high density of UDN has significantly decreased compared to its performance in a low-density environment. This is due to the increased interference from multiple gNBs as well as the high number of handovers in a high density of UDN. These tables provide a useful mapping to determine the optimal TTT values for different scenarios.

\begin{figure*}[htbp]
	\centering
	\subfigure[Handover rate for different velocities and TTT values ]{
		\includegraphics[width=0.45\textwidth]{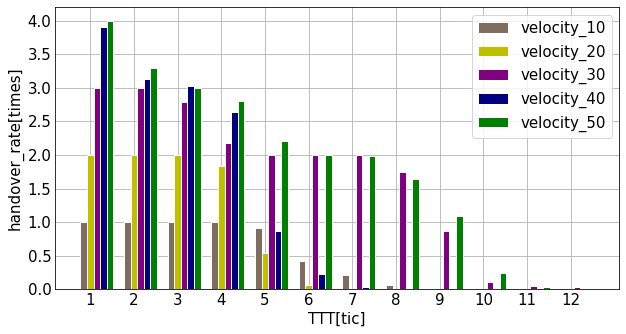}}
	\label{fig.sub.3}
	\subfigure[Average SINR for different velocities and TTT values]{
		\includegraphics[width=0.45\textwidth]{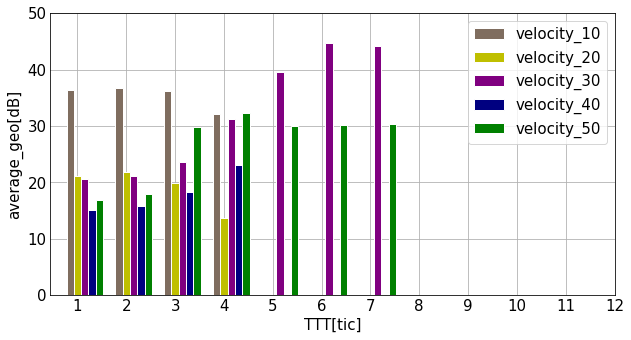}}
	\label{fig.sub.4}
    \caption{Simulation results for different velocities and TTT values for Case B}
    \label{velocity}
\end{figure*}

\begin{table}[htbp]
\caption{Average SINR of TU for Case B}
\begin{center}
\begin{tabular}{ |p{2.8cm}|p{0.5cm}|p{0.5cm}|p{0.5cm}|p{0.5cm}|p{0.5cm}|  }
\hline
\backslashbox[32mm]{\textbf{TTT}}{\textbf{ho\_avg\_sinr}}{\textbf{density}}
&\makebox{10}&\makebox{20}&\makebox{30}
&\makebox{40}&\makebox{50} \\
 \hline
    1 & 16.93 & 4.05 & 4.11  & 5.2    & 4.04\\
 \hline
    2 & 17.93 & 6.50 & 5.86  & 6.38   & 5.45\\
 \hline 
    3 & 29.73 & 11.78 & 10.36 & 10.11  & 9.82\\
 \hline 
    4 & 32.28 & 19.22  & nan   & nan    & nan\\
 \hline
    5 & 29.93 & 26.88 & nan   & nan    & nan\\
 \hline
    6 & 30.14 & 29.46 & nan   & nan    & nan\\
 \hline
    7 & 30.36 & nan & nan   & nan    & nan\\
 \hline
    8 & nan   & nan & nan   & nan    & nan\\
 \hline 
    9 & nan   & nan   & nan   & nan    & nan \\
 \hline
    10 & nan  & nan   & nan   & nan    & nan\\
 \hline
    11 & nan  & nan   & nan   & nan    & nan\\
 \hline
    12 & nan  & nan   & nan   & nan    & nan\\
 \hline
\end{tabular}
\end{center}
\label{caseB}
\end{table}

\subsection{Effect of variable TTT values and velocities with a fixed density of UDN on handover}

The simulation results in Fig. \ref{velocity} provide insight into the relationship between TTT values, velocities, and handover performance in a fixed density of UDN (20) scenario. These results can serve as a reference for determining the optimal TTT values for different use cases.
In Fig. \ref{velocity}(a), it is observed that as the TTT value increases, the handover rate decreases with the increase in velocity. This is because, with the increase in TTT value, the TU's connection with the network is maintained for a longer period, leading to fewer handovers. However, this does not necessarily translate to better handover performance, as can be seen in Fig. \ref{velocity}(b). When TTT is 1, the average SINR decreases from 36.29 dB at 10 km/h to 16.93 dB at 50 km/h but the handover rate increase from 1 at 10km/h to 4 at 50 km/h. This is due to the fast-moving TU encountering more frequent handovers, leading to increased signal interference and decreased handover performance. It can be observed that as the velocity of the TU is 50 km/h, the average SINR rises from 16.93 dB at $TTT = 1$ to 30.3 dB at $TTT =7$. When TTT is between 4 and 8 with the optimal velocity of 30 km/h, the TU can achieve the best handover performance.
\section{Conclusion}
This paper has conducted an analysis of the impact of TTT values, UDN densities, and TU velocities on 5G handover performance. The simulation results show that the TTT value plays a crucial role in determining the handover rate, and finding a proper balance between the TTT value, UDN density, and TU velocity is crucial for optimizing handover performance. The authors have also developed a simulation tool in Python to evaluate the handover times and performance for different scenarios. In future work, the authors aim to improve handover performance by using machine learning algorithms in 5G/6G wireless networks.

\section{acknowledgement}
This work has been funded by the German Federal Ministry of Education and Research as part of the AI4mobile project, with a funding number of 16KIS1170K. The authors acknowledge the contributions of all AI4Mobile partners, but the content of the paper is the sole responsibility of the authors and may not necessarily reflect the views of the project as a whole.

\bibliographystyle{IEEEtran}
\bibliography{references}

\end{document}